\documentclass[useAMS,usenatbib]{mn2e}

\usepackage{epsfig}

\def\msun{M$_\odot$}
\def\lesssim{\mathrel{\hbox{\rlap{\hbox{\lower3pt\hbox{$\sim$}}}\hbox{\raise1pt\hbox{$<$}}}}}
\def\gtrsim{\mathrel{\hbox{\rlap{\hbox{\lower3pt\hbox{$\sim$}}}\hbox{\raise1pt\hbox{$>$}}}}}

\newcommand\aj{\rmfamily{AJ}}%
\newcommand\araa{\rmfamily{ARA\&A}}%
\newcommand\apj{\rmfamily{ApJ}}%
\newcommand\apjl{\rmfamily{ApJ}}%
\newcommand\apjs{\rmfamily{ApJS}}%
\newcommand\aap{\rmfamily{A\&A}}%
\newcommand\aaps{\rmfamily{A\&AS}}%
\newcommand\mnras{\rmfamily{MNRAS}}%
\newcommand\pasp{\rmfamily{PASP}}%
\newcommand\nat{\rmfamily{Nature}}%
\newcommand\iaucirc{\rmfamily{IAU~Circ.}}%
\title[Location Diagnostic of Peculiar Supernovae]{Locations of Peculiar Supernovae as a Diagnostic of Their Origins}
\author[F.~Yuan et al]
{\parbox{\textwidth}{
Fang~Yuan,$^{1,2}$\thanks{E-mail:
yuanfang@mso.anu.edu.au}
Chiaki~Kobayashi,$^{3}$
Brian~P.~Schmidt,$^{1,2}$
Philipp~Podsiadlowski,$^{4}$
Stuart~A.~Sim$^{1,2}$ and Richard~A.~Scalzo$^{1,2}$
}\vspace{0.4cm}\\
$^{1}$Research School of Astronomy and Astrophysics, Australian National University, Canberra, ACT 2611, Australia\\
$^{2}$ARC Centre of Excellence for All-sky Astrophysics (CAASTRO)\\
$^{3}$School of Physics, Astronomy and Mathematics, University of Hertfordshire, Hatfield, AL10 9AB, UK\\
$^{3}$Department of Astrophysics, Oxford University, Oxford OX1 3RH, UK}

\begin{document}

\date{Accepted -  Received -}

\pagerange{\pageref{firstpage}--\pageref{lastpage}} \pubyear{2013}

\maketitle

\label{firstpage}

\begin{abstract}
We put constraints on the properties of the progenitors of peculiar Òcalcium-richÓ transients using the distribution of locations within their host galaxies.
We confirm that this class of transients do not follow
the galaxy stellar mass profile and are more likely to be found in remote
locations of their apparent hosts. 
We test the hypothesis that these
transients are from low metallicity progenitors by comparing their spatial
distributions with the predictions
of self-consistent cosmological simulations that include star formation and chemical enrichment. We find that
while metal-poor stars and our transient sample show a consistent preference
for large offsets, metallicity alone cannot explain the extreme
cases. Invoking a lower age limit on the progenitor helps to improve the
match, indicating these events may result from a very old metal-poor
population. We also investigate the radial distribution of globular cluster systems, and show that they too are consistent with the class of calcium-rich transients. 
Because photometric upper limits exist for globular clusters for some members of the class, a production mechanism related to the dense environment of globular clusters is not favoured for the calcium-rich events. 
However the methods developed in this paper may be used in the future to constrain the effects of low metallicity on radially distant core-collapse events or help establish a correlation with globular clusters for other classes of peculiar explosions.
\end{abstract}


\section{Introduction}

Supernovae (SNe) are luminous events caused by the explosions of stars.
They are divided into observational subclasses according to line features
from different chemical elements appearing in the spectra;
type Ia supernovae (SNe~Ia) are generally
attributed to the thermonuclear disruption of a white dwarf
\citep[see e.g.,][]{2000ARA&A..38..191H}, while type
Ib/c supernovae (SNe~Ib/c) and type II supernovae (SNe~II) result from the
gravitational collapse of the core of a massive star at the end of its life
\citep[main sequence mass $\gtrsim$8~\msun,][]{2009ARA&A..47...63S}.
However, details of the progenitor systems, the physics of the explosions,
and the evolutionary paths leading to those explosions are still unclear and under
investigation.

Automated wide field optical transient surveys, e.g., the Texas Supernova
Search \citep[TSS,][]{quimby_thesis}/ROTSE Supernova Verification Project
\citep[RSVP,][]{yuan_thesis}, the Catalina Real-Time Transient Survey
\citep[CRTS,][]{2009ApJ...696..870D}, the Palomar Transient Factory
\citep[PTF,][]{2009PASP..121.1334R} and the Panoramic Survey Telescope
and Rapid Response System \citep[Pan-STARRS,][]{2002SPIE.4836..154K},
have in recent years not only discovered hundreds of new SNe of all types,
but discovered new types of transient events in hitherto unexplored regions
of parameter space. 

An increasing number of unusual events are being discovered by these surveys, 
often occurring in host environments poorly sampled by previous supernova
surveys, which targeted nearby massive galaxies.  For example, PTF10ops is a peculiar
SN~Ia $\sim$148 kpc away from the nearest galaxy \citep{2011MNRAS.418..747M}.
It is puzzling to also find core-collapse explosions at remote locations,
given the expected low local star-formation rate.
The massive star progenitors of these core-collapse events either have very low metallicity
\citep[e.g., SN 1983K,][]{1985ApJ...289...52N, 1990PASP..102..299P}
or have gained very high velocity and migrated outward from the galaxy centre
\citep{2011A&A...536A.103Z}.
Alternatively, a core-collapse could be delayed if it is the result of
the merger of two white dwarfs \citep[e.g., an ONeMg and a He white dwarf; cf.][]{2007MNRAS.380..933Y}.
At present, it is not clear whether such mergers prefer old populations
and/or low-metallicity environments.

\citet{2010Natur.465..322P} identified a group of faint events,
spectroscopically similar to type Ib/c SNe and displaying prominent calcium
lines in the nebular phase.  Whether these events constitute a homogeneous
class with a common physical explosion mechanism remains unclear, as does 
the nature of that mechanism (thermonuclear vs. core-collapse).
A core-collapse scenario has been put forward for one of these events, SN
2005cz, in which an 8 to 12~\msun~progenitor lost its hydrogen-rich envelope
through binary interaction \citep{2010Natur.465..326K}. However, detailed
studies at and around the supernova location do not reveal evidence of a young
stellar population \citep{2011ApJ...728L..36P}.  The core-collapse scenario
for these transients also struggles to explain the large fraction found in
early-type host galaxies.  An alternative explanation for the small amount
of $^{56}$Ni synthesised in these explosions is a scenario involving a helium
detonation on the surface of an accreting white dwarf
\citep{2007ApJ...662L..95B, 2010Natur.465..322P, 2010ApJ...715..767S, 2011ApJ...738...21W, 2012MNRAS.420.3003S}. 
Both the core-collapse and the thermonuclear models can reproduce qualitatively a large Ca/O line ratio \citep{2010Natur.465..326K, 2011ApJ...738...21W}, but comprehensive spectral series modelling across the range of observed calcium-rich transients is still lacking.
 
Several more events were discovered by PTF that share some common properties
with this calcium-rich group. In particular, \citet{2012ApJ...755..161K}
identified a class of five objects
\citep[including two events from][]{2010Natur.465..322P} using both photometric and spectroscopic criteria. Interestingly, all these peculiar PTF transients are found at large distances from their host galaxies \citep{2011ApJ...732..118S, 2011MNRAS.418..747M}. 

No single model so far has successfully reproduced all the observed features of the peculiar ``calcium-rich'' transients. 
In this paper, we take an alternative, model-independent approach to constrain the properties of their progenitors by statistically comparing their locations to the overall
stellar population in their host galaxies. We define our sample selection in
Section~\ref{section:sample}. We quantitatively evaluate the transient
locations relative to stellar mass distribution in
Section~\ref{section:light}. We compare the observed distribution with theoretical predictions of cosmological simulations in
Section~\ref{section:metallicity} and investigate the expected radial distribution of objects associated with globular clusters in Section~\ref{section:gc}. 
The implications are discussed in Section~\ref{section:discussion} and we conclude
in Section~\ref{section:conclusion}.

\section{Sample of Transients}\label{section:sample}
We first consider the collection of five transients selected based on both
photometric and spectroscopic properties by \citet{2012ApJ...755..161K}. This
will be referred to as the K12 sample. This sample includes two events
(SN 2005E and SN 2007ke) from \citet{2010Natur.465..322P}, as well as
three events selected from a sample of over a thousand transients discovered
by PTF in its first two years of operation:
PTF09dav, PTF10iuv (aka SN 2010et), and PTF11bij.
Compared to typical supernovae, the defining
characteristics are a moderately faint peak luminosity (absolute R magnitude
of $\sim-16$), relatively fast rise/decay ($\sim$12~-~15 days), normal
photospheric velocity, a quick spectral transition into the nebular phase, and
nebular spectra dominated by calcium emission. 

Another five events (SN 2000ds, 2001co, 2003H, 2003dg and 2003dr) have been 
spectroscopically selected as peculiar calcium-rich Type Ib/c SNe
\citep{2003IAUC.8159....2F,2010Natur.465..322P}. Their photospheric spectra
resemble that of typical SNe Ib/c, lacking silicon, hydrogen and sometimes
helium lines. Their nebular spectra are dominated by calcium emission. While
it is impossible to verify the photometric properties for these objects, there
is evidence that they are sub-luminous compared to normal SNe
\citep{2010Natur.465..322P}, and consistent with the
objects in the K12 sample. We therefore include them in the ``calcium-rich''
group to maximise our sample size for statistical study.

All seven events from  \citet{2010Natur.465..322P} were discovered/observed by the Lick Observatory Supernova Search \citep[LOSS,][]{2001ASPC..246..121F}. We regard them as a reasonably complete and uniform set, although we will discuss possible selection effects in both the PTF and LOSS samples in Section~\ref{section:bias}. Observations of SN 2005cz, a single event discovered by another group \citep{2010Natur.465..326K}, show it to be a convincing member of the ``calcium-rich'' class. It is, however, excluded in our study to avoid further complicating the understanding of its selection bias.

We follow \citet{2012ApJ...755..161K} in identifying the host as the nearest galaxy in projected distance at a redshift consistent with the transient. The measured distance between a transient and the centre of its host is thus a lower limit on the real distance.

\section{Locations Compared to Stellar Mass}\label{section:light}
In this section we compare the distribution of the ``calcium-rich'' transients with the stellar mass distribution of their host galaxies. We use $K$-band photometry from  the Two Micron All Sky Survey (2MASS) catalog \citep{2006AJ....131.1163S} to trace the overall stellar mass. 

We analyse images from the 2MASS All-Sky Data Release. For the host of
SN 2005E and SN 2007ke, several overlapping images must be stitched together
to get good coverage of the target galaxies.
We subtract bright stars near the galaxies using the IRAF\footnote{IRAF is distributed by the National Optical Astronomy Observatories, which are operated by the Association of Universities for Research in Astronomy, Inc., under cooperative agreement with the National Science Foundation.} DAOPHOT package.
We extract the moments of the galaxy isophotes using SExtractor
\citep{1996A&AS..117..393B}, then use the IRAF package ELLIPSE to extract
fluxes in elliptical annuli, with the semi-major axis increased in one-pixel
increments and a common centroid, ellipticity, and position angle for all
annuli fixed at the values obtained from SExtractor.
We fit the smooth part of the surface brightness profile between one-half and
one Kron radius \citep[2.5 times the first moment
radius,][]{1980ApJS...43..305K} with a S$\acute{e}$rsic model
\citep{1968adga.book.....S}, but allow the S$\acute{e}$rsic index to vary
freely. We integrate the resulting surface brightness profile to estimate
the total flux of the galaxy. Fractions of stellar light enclosed by
elliptical apertures defined by the transient locations are calculated
and listed in Table~\ref{tab:location}.

\begin{table*}
\centering
\scriptsize
\begin{minipage}{\textwidth}
\caption{Locations of calcium-rich transients in their host galaxies}
\label{tab:location}
\begin{tabular}{@{}cccccccc@{}}
\hline
Transient & Host & Host & Host & Host r$_{\rm eff}$\footnote{Major axis of the ellipse enclosing half of the galaxy $K$-band light.} &  Distance from &  Fraction of Mass &  Fraction of Mass \\
 Name &  & Type\footnote{Classification from HyperLeda \citep{2003A&A...412...45P}.} &  M$_{K}$ &  (kpc) &  Host (kpc) &  Enclosed\footnote{Estimated in elliptical apertures defined by the transient locations.} 
 &  Enclosed\footnote{Estimated in circular apertures defined by the transient locations.}\\

\hline
SN 2000ds & NGC 2768 & E & $-24.8 (-0.1, +0.2)$ & 6.2 & 3.3 & $0.58 (-0.05, +0.11)$ & $0.43 (-0.04, +0.08)$\\
SN 2001co & NGC 5559 & SBb & $-24.2 (-0.1, +0.0)$ & 5.6 & 6.6 & $0.70 (-0.08, +0.02)$ & $0.77 (-0.08, +0.03)$\\
SN 2003H  & IC 2163 & Sc & $-24.6 (-0.1, +0.4)$ & 6.6 & 6.7 & $0.43 (-0.04, +0.17)$ & $0.61 (-0.06, +0.23)$\\
SN 2003dg & UGC 6934 & Sc  & $-23.0 (-0.2, +0.0)$ & 4.4 & 1.4 & $0.11 (-0.02,+0.00)$ & $0.26 (-0.05, +0.00)$\\
SN 2003dr & NGC 5714 & Sc  & $-22.7 (-0.1, +0.0)$ & 3.8 & 2.5 & $1.00 (-0.10, +0.00)$ & $0.55 (-0.06, +0.02)$\\
SN 2005E  & NGC 1032 & S0/a  & $-24.3 (-0.2, +0.0)$ & 3.6 & 23.9 & $1.00 (-0.04, +0.00)$ & - \\ 
SN 2007ke & MCG +07-07-003 & E/S0 & $-23.1(-0.1, +0.0)$ & 0.9 & 8.2 & $1.00 (-0.05, +0.00)$ & - \\ 
PTF09dav & 2MASX J22465295+2138221 & S? & $-23.0 (-0.3, +0.0)$ & 5.5 & 40.7 & $1.00 (-0.03, +0.00)$ & - \\ 
PTF10iuv & CGCG 170-011 & S? & $-23.4 (-0.2, +0.0)$ & 2.6 & 35.6 & $1.00 (-0.02, +0.00)$ & - \\
PTF11bij & IC 3956 & S? & $-24.1 (-0.1, +0.0)$ & 3.1 & 34.0 & $1.00 (-0.04, +0.00)$ & - \\ 
\hline
\end{tabular}
\end{minipage}
\end{table*}

SN 2003H lies between a pair of interacting galaxies. We ignore the brightening on the outskirts of IC 2163 due to contamination from NGC 2207. Nevertheless, it is likely that the local environment has been affected by the interaction and therefore cannot be fully represented by its distance to either galaxy in the pair. 

For SN 2007ke, both the transient and the identified host MCG +07-07-003 lie
in the bright galaxy NGC 1129. We remove the background light from NGC 1129 by
modelling its surface brightness profile in the same way as described above.
If NGC 1129 is the host galaxy, the location of SN 2007ke will not be as
extreme (within $\sim$50\% of light), but uncertainty in this single event
does not change our main conclusions.

If the surface brightness in the host galaxy drops faster (slower) than our
model in the outer regions of the galaxy, the fraction of light enclosed by
the elliptical isophote passing through the transient's location will be
larger (smaller) than our calculation. We estimate the systematic uncertainty
in the surface brightness by considering profiles with S$\acute{e}$rsic
indices of 0.2 (steep profile) and 4 (flat profile) outside the fitted region.

\begin{figure}
\includegraphics[width=82mm]{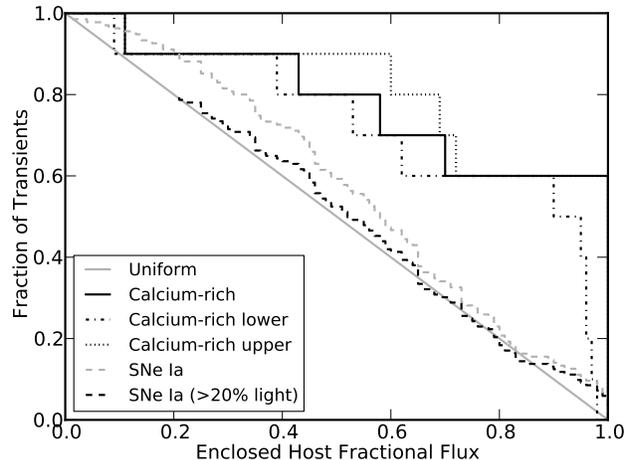}
\caption{Distribution of ``calcium-rich'' transients vs. enclosed fractional galaxy $K$-band light. The upper and lower limits are derived by fixing the galaxy S$\acute{e}$rsic index to 0.2 and 4 respectively (see text). The grey line shows the expectation if the transients follow the light distribution. The grey and dark dashed histograms represent respectively SNe Ia and SNe Ia found outside 20\% $K$-band light.}
\label{fig:dist_mass}
\end{figure}

Figure~\ref{fig:dist_mass} shows that the ``calcium-rich'' transients do not
follow the distribution of $K$-band light, a proxy for the stellar mass, in
their host galaxies. Instead, they have a strong preference for remote
locations.  A Kolmogorov-Smirnov (K-S) test shows that even the limiting case
of a flat profile (S$\acute{e}$rsic index of 4) is inconsistent at greater
than 99\% confidence with a scenario in which the transient locations
uniformly track the $K$-band flux.

For comparison, we also estimate the fraction of $K$-band light enclosed by
isophotes passing through the locations of 135 SNe Ia in early-type hosts
discovered between January 1990 and September 2011.
The SN types, coordinates and host galaxy identifications were obtained from
the IAU Central Bureau for Astronomical Telegrams (CBAT)
website\footnote{http://www.cbat.eps.harvard.edu/lists/Supernovae.html}. A SN
Ia is included in the sample if its host galaxy has a morphology type of E, S0
or SA0 as classified in the NASA/IPAC Extragalactic Database
(NED)\footnote{http://ned.ipac.caltech.edu/} and has a redshift below 0.07. Our full SN~Ia sample shows a
deficit in the inner region of galaxies. To determine whether this reflects a
true deficiency in the underlying distribution, we need to know the search
efficiency as a function of projected galactocentric distance, especially for
small distances.  This factor is hard to determine precisely, since our
sample is not drawn from a single survey with a well-understood selection
function.  We can proceed by making the reasonable assumption that the
discovery efficiency drops towards the bright core of galaxy
due to low contrast or blending, but is relatively flat outside 20\% of
stellar light. If we only include SNe Ia found outside 20\% of stellar light
(and assume 20\% of events are cut by this criteria), the remaining SNe~Ia
track the light well, as seen in Figure~\ref{fig:dist_mass}. This agrees with
the result found in $r$-band by \citet{2008MNRAS.388L..74F}, although $K$-band is a better tracer of mass than $r$-band \citep{1993ApJ...418..123R}. The decreased
efficiency near galaxy centres has a minimal impact on our results regarding
the ``calcium-rich'' sample.

Using a different fractional flux measurement, \citet{2008ApJ...687.1201K}
have shown that SNe Ia and SNe II follow the distribution of $g^\prime$-band
light while SN Ic and long GRBs are concentrated in the brightest regions of
their hosts \citep{2006Natur.441..463F}. The ``calcium-rich'' transients
appear to be distributed differently with respect to host galaxy light from
any of these samples.

\section{Metallicity and Age Constraints}\label{section:metallicity}
We now translate the distribution of the ``calcium-rich'' transients with
respect to host galaxy surface brightness into constraints on the age and
metallicity of their progenitors.

Galaxies are complex systems, often containing multiple stellar populations
each with a different age and metallicity.  Central or average metallicities
of host galaxies can be quite different from local metallicities at the
supernova sites.
Attempting to extrapolate the metallicity from the bright centre of a galaxy
to its outskirts, where most of our events are located, is also dangerous
since there is a large scatter in the gradient-mass relation
\citep{2010MNRAS.408..272S}.  If the progenitors are very old, they may also
have traveled far from the original formation site by the time of explosion,
so the local metallicity may not give an accurate estimate of the progenitor metallicity.
Finally, there is a significant scatter in the metallicity distribution
function at a given radius, i.e., there are some metal-poor stars in the
centre and metal-rich stars in the outskirts of galaxies.
To take all of these effects into account, we statistically compare the
distribution of projected galactocentric distances in our sample with the
predicted distribution of stars with known metallicity 
in self-consistent cosmological simulations.

The simulations used in this comparison include the relevant physics such as
hydrodynamics, star formation, supernova feedback, and chemical enrichment,
the details of which are described in \citet{2007MNRAS.376.1465K}. Here we use
a new simulation run with a resolution of $5 \times 10^6 M_\odot$, WMAP-5
cosmology and the Kroupa initial mass function \citep{2001MNRAS.322..231K}.
A star formation parameter $c = 0.02$ and number of feedback neighbours
($N_\mathrm{FB} = 576$)
are used in order to match the observed cosmic star formation rate
history and the mass-metallicity relations of galaxies.
Metallicities, elemental
abundances, ages, and positions of stars are recorded for analysis.
Collections of particles in the simulation are identified as galaxies
with the friend-of-friends method. 

We define metallicity as the logarithm of the mass fraction of elements heavier than helium compared to that of the Sun ([M/H]). 
From the simulations, we calculate the metallicity distributions as a function
of projected enclosed stellar mass within host galaxies. We use circular
apertures, since the mass distributions are approximately spherical in the
outer regions of galaxies.  We integrate the stellar masses along the
line-of-sight since the observational data are measured in projected radii.
To be consistent with the simulations, we remeasure the enclosed $K$-band flux of
our peculiar events in circular apertures.
In cases where the transient is too far from the host to accurately determine
the enclosed flux, we use the flux ratios from the previous section
(see Table~\ref{tab:location}).  We expect the choice of aperture shape
to have a minimal effect on our results.

The observed $K$-band total luminosities ($L_K$) of the host galaxies are converted to stellar masses for comparison with simulations. $M/L_K$ values of 0.8, 0.7 and 0.5 are used for E/S0, S0a/b and Sbc/d type galaxies respectively \citep{2001ApJ...550..212B} and $M/L_K$ of 0.5 is used for the three hosts of unknown type. The resulting mass range is about $1.2\times10^{10}$~\msun~to $1.4\times10^{11}$~\msun, and therefore we select the galaxies in the simulation whose total stellar masses fall in the range $1\times10^{10}$~\msun~to $1\times10^{11}$~\msun. Note that the result does not change qualitatively if a wider range of galaxies is included, as the stellar mass distribution peaks within the selected host mass range. 

\begin{figure}
\includegraphics[width=82mm]{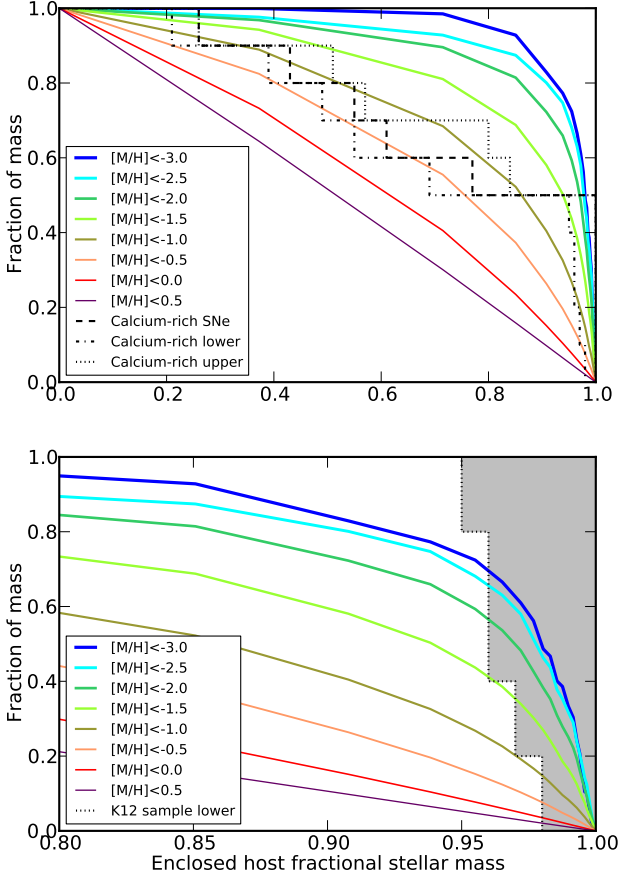}
\caption{Observed distribution of the peculiar transients compared to those of
stars of different metallicity within galaxies as predicted by our simulation.
The full sample is shown in the top panel. In the bottom panel, the shaded
area represents the contour for the K12 sample. Note that in the bottom panel,
the x-axis range is zoomed in by a factor of 5 to enable the reader to
distinguish visually between curves corresponding to different models.}
\label{fig:dist_metal}
\end{figure}

Figure~\ref{fig:dist_metal} shows the distribution of  ``calcium-rich'' transients compared to the stellar distribution for different metallicity ranges inferred from our simulations. In general, the ``calcium-rich'' transients lie further away from the host than stars of [M/H] $\lesssim-0.5$, consistent with a relatively low metallicity origin.

If only the five events in the K12 sample are considered, the distribution is
skewed further from the host and is completely different than any of our
stellar metallicity curves in Figure~\ref{fig:dist_metal}. One way to resolve
the discrepancy is to introduce an age limit on the stellar population. In the
CDM picture, old stars form in small galaxies, which successively merge to
form a large galaxy. Old stars that formed in a deep potential well tend to
be located in the galactic bulge at present, and the chemical enrichment
timescale is so short that the metallicities of the majority of such old stars
are not very low. On the other hand, old stars that formed in a shallow
potential well, whose parent galaxy/cluster is disrupted, tend to be located
in the outer halo at present, where no young stars form.

\begin{figure}
\includegraphics[width=82mm]{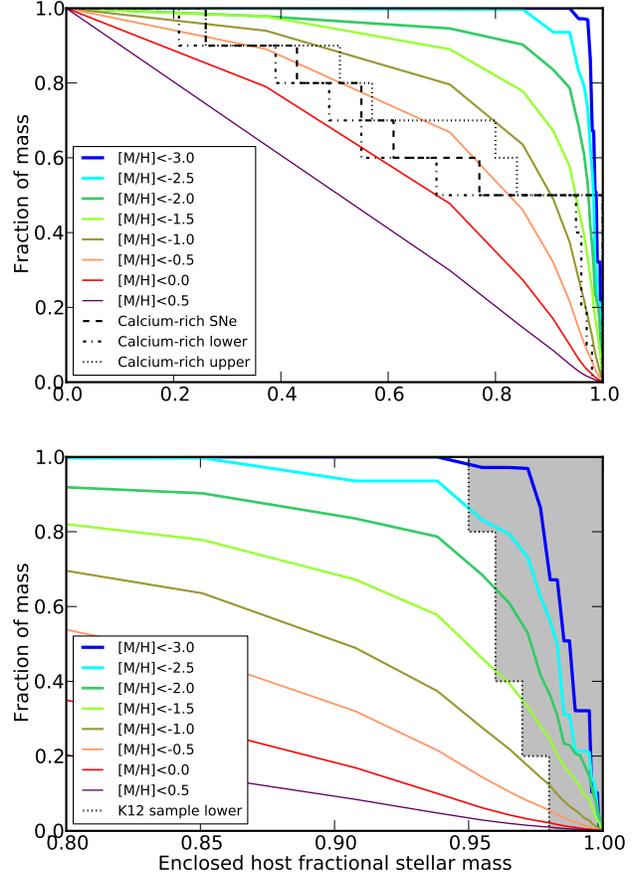}
\caption{Observed distribution of the peculiar transients compared to those of
stars of different metallicity and older than 10 Gyr within galaxies as
predicted by our simulation. The full sample is shown in the top panel. In the
bottom panel, the shaded area represents the contour for the K12 sample.  As
in Figure \ref{fig:dist_metal}, in the bottom panel the x-axis range is zoomed
in by a factor of 5 to enable the reader to distinguish visually between
curves corresponding to different models.}
\label{fig:dist_metal_old}
\end{figure}

Our sample does not have the statistical power to constrain the age. For illustrative purpose, we set a cutoff age of 10 Gyr, corresponding to the lifetime of a star with zero main-sequence mass of about 1~\msun. If we only trace stars that are older than 10 Gyr in our simulation, a dominant fraction of the metal poor population is found at large distances from the galaxy centres. As can be seen in the bottom panel of Figure~\ref{fig:dist_metal_old}, even at this old age, the curve that comes close to the K12 sample has extremely low metallicity ([M/H] $\lesssim-3$).

\section{Correlation with Globular Clusters} \label{section:gc}
Another possibility leading to such large offset for classes of objects is if they are associated with globular clusters (GCs). GCs are detected throughout the extended
halos of galaxies \citep[sometimes over 100~kpc away from the
centre,][]{2006ARA&A..44..193B}. 
 The high stellar densities in GCs lead to an
enhanced rate of close binaries and unusual objects, which could provide unique
channels for exotic transients. 
In this section, we test whether the progenitor population could be correlated
with GC systems.

Direct photometric observations have failed to detect GCs of typical brightness at the sites of some of our nearby events. For SN 2000ds, Hubble Space Telescope imaging revealed no progenitor brighter than $M_{V}\sim-5.5$ \citep{2003PASP..115....1V}, ruling out an association with a typical GC at $M_{V}\sim-7$. For SN 2005cz, a likely ``calcium-rich'' member that is not included in our statistical study, a non-detection down to $M_{V}\sim-6.9$ was obtained \citep{2011ApJ...728L..36P}. This turns into a fairly restrictive limit, particularly considering that brighter GCs tend to be more massive and more centrally concentrated \citep[cf.][]{2000ApJ...539..618M} and thus more likely to harbour unusual objects. Because the calcium-rich class is still poorly defined, and there exists potentially other large-offset transient populations (e.g., PTF10ops) we, despite these photometric limits, carry out the following comparison. This analysis should be useful for further studies where observations of comparable absolute depth are not achievable for more distant supernovae. The method described can also be generalised and applied to statistical studies of other transient classes which currently have too few members for study.

For each host, we attempt to model the GC profile and calculate the fraction
of GCs within (or outside) the transient location, as we did for the stellar
light in section~\ref{section:light}. The projected radial
distribution of globular clusters in a galaxy is often described by a
power law, with slopes between $\sim-2.5$ and $\sim-1.5$ \citep[][and
references therein]{2006ARA&A..44..193B}. More luminous galaxies tend to have
shallower profiles while no obvious trend is found with galaxy type
\citep{1986AJ.....91..822H}. This luminosity-slope relation is at least partly
related to the fact that brighter galaxies are often larger in their physical
extent.

\begin{figure}
\includegraphics[width=82mm]{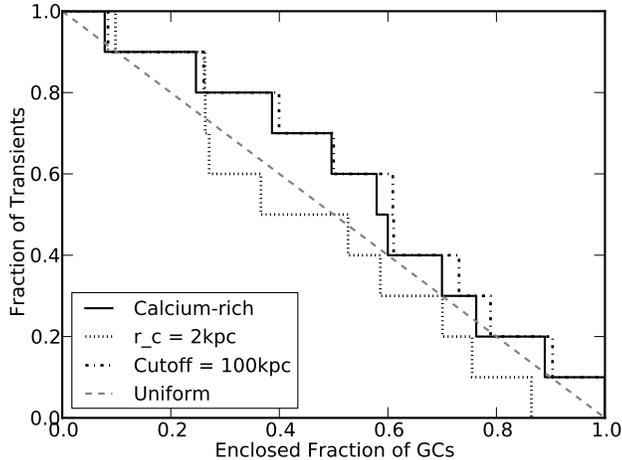}
\caption{Cumulative distribution of ``calcium-rich'' transients as a function of enclosed fraction of GCs. The dashed line shows the expectation if the transients follow the GC distribution. The dotted line is for a fixed GC profile core size of 2~kpc and the dash-dotted line is for a cutoff of GC extent at 100~kpc.}
\label{fig:gc_radial}
\end{figure}

Since the GC radial distribution tends to flatten at the centre of a galaxy,
we describe the GC projected surface density at radius $r$ as
$\rho_{0}(1+(r/r_{c})^{-\alpha})^{-1}$. The core size $r_{c}$ is derived using
the relation illustrated in Figure.~10 of \citet{1996ApJ...467..126F}. The
slope $\alpha$ is estimated with Eq.~(11) in \citet{1986AJ.....91..822H}.
Integrated $V$-band magnitudes of the galaxies are obtained from the NED
database or converted from SDSS g and r model magnitudes. For three hosts
without $V$-band or SDSS photometry, we assume $(V-K)=3.2$ mag (the median for
the other seven galaxies). We also apply a realistic but somewhat arbitrary
cutoff for the GC population extent of 120~kpc. The projected distance
between a transient and the centre of its apparent host is derived using the
angular separation and the CMB rest-frame distance in the NED database,
assuming H$_{\rm 0}$=70~km~s$^{-1}$ Mpc$^{-1}$, $\Omega_{\rm m}$=0.27 and
$\Omega_{\Lambda}$=0.73. The resulting distribution is shown in
Figure~\ref{fig:gc_radial}. While the histogram seems to suggest the
``calcium-rich'' transients may have a slightly flatter distribution than the
GCs at the inner regions of the galaxies, the discrepancy is not statistically
significant. We show that the result is not very sensitive to the choice of
cutoff distance as all of our transients lie within 50~kpc. On the other hand,
we see a stronger dependence on the core sizes. Given the many uncertainties
in our GC profile estimate, we can only conclude that we
do not rule out the possibility that ``calcium-rich'' transients originate in
GCs.

The number of GCs per unit stellar mass increases with decreasing galaxy mass
at low masses \citep{2007ApJ...670.1074M}. This trend, combined with the
galaxy stellar mass function \citep[e.g., from][]{2012MNRAS.421..621B}, means that a
significant fraction of GCs exist in relatively faint/small galaxies.
Unfortunately, seven of our 10 SNe were discovered in the LOSS survey,
which targets large nearby galaxies and which may under-represent low surface
brightness hosts \citep{2011MNRAS.412.1419L}. The
tight distribution of $K$-band magnitudes of the identified hosts roughly agree
with the typical luminosity of a LOSS galaxy. For a GC-related sample
collected entirely through untargeted surveys in future, we would expect to
see a wider range of host masses/luminosities.

\section{Discussion}\label{section:discussion}

\subsection{Sample Bias}\label{section:bias}
Although the PTF ``calcium-rich'' transients all occurred far from the centres
of their respective host galaxies, this is not a selection criterion for the
class.  The survey strategy can potentially introduce a bias, but PTF
regularly discovers SNe at a range of galactocentric distances, suggesting
that selection bias is not primarily responsible for the observed locations
of the ``calcium-rich'' transients.  On the other hand, the targeted LOSS
survey imposes a bias in their sample against objects lying far away from any
bright host. By simply combining events from PTF and LOSS, we run the risk of
mixing two samples drawn from different populations.  A larger sample drawn
from ongoing and future untargeted surveys will improve the situation.
For now, we try to verify that each of our sub-samples is
statistically consistent with the overall population.

The LOSS survey monitors a predefined nearby galaxy population with the 0.76-m
Katzman Automatic Imaging Telescope (KAIT) \citep{2001ASPC..246..121F}. We
estimate the equivalent scale radii covered by KAIT's relatively small
($6.7\times6.7~\mathrm{arcmin}^{2}$) field of view at the distances of the
seven LOSS discoveries. For the 2 (4) nearest events, this value is less than
30 (40) kpc.  KAIT may therefore have missed transients with separations from
their host galaxy larger than this.

If the 3 events from PTF are drawn from the distribution of all ten
``calcium-rich'' transients, we expect another 1 (2) events detected inside
about 50\% (70\%) stellar mass. Recognising a low luminosity ``gap'' transient
near the bright core of a host galaxy can be challenging. For example in the
host of PTF10iuv (r~=~14.2), a 19th magnitude object would be $\sim$25\%
brighter than the underlying host at around 70\% stellar mass. However,
the number of transients potentially missed is still roughly consistent with
fluctuations from Poisson statistics given the small sample size.

\subsection{Implications} \label{section:implications}
A metallicity effect remains an appealing explanation for the large offsets of the ``calcium-rich'' transients. In either a thermonuclear or a core-collapse model, metallicity can affect the evolution of a progenitor star, impacting on its age, structure and chemical composition at the time of explosion. For example, production of a normal SN Ia (in a single-degenerate scenario) is inhibited at low metallicity, if the wind from the accreting white dwarf is not strong enough to stabilise the mass transfer \citep{1998ApJ...503L.155K}. Perhaps an exotic sub-luminous explosion can occur under these circumstances. In the case of a massive star, wind-driven mass loss is closely related to metallicity. At lower metallicity, the mass loss becomes weaker and a larger helium core is retained, which may form a black hole directly, resulting in a faint supernova \citep{2011ApJ...730L..14K} or no explosion \citep{2003ApJ...591..288H}.

Moreover, the extreme cases of ``calcium-rich'' transients (the K12 sample)
seem to lie further away from the host than the region where metallicity falls
steeply. An additional parameter may be needed to interpret this distribution.
While we cannot strongly
constrain the age with our sample, our simulation show that extremely metal
poor stars older than 10 Gyr are expected to be predominantly found at  the
same distant locations in galaxies as the K12 transients. This trend also
argues strongly against any massive star origin. This conclusion is consistent
with the fact that some of the hosts are early-type galaxies without evidence for recent star formation.

We also investigated if these peculiar transients could be associated with globular cluster systems, which often have a more extended spatial distribution than stellar light. In section~\ref{section:gc} we showed that the radial distribution of the ``calcium-rich'' transients is consistent with the GC distributions. However, deep photometric observations argue otherwise and the discussed sample is not likely to originate from GCs. The GC connection may be more relevant for transients with suspected binary progenitors \citep{2009ApJ...695L.111P}, such as PTF10ops \citep{2011MNRAS.418..747M}. One possible interpretation for this unusual subluminous SN Ia is that it arises from the merger of two equal-mass white dwarfs \citep{2010Natur.463...61P}. Sufficiently violent mergers might preferentially occur in the dynamical environment of globular clusters where non-sychronized and/or eccentric binaries could be more common. Alternatively, \citet{2009ApJ...705L.128R} have shown that the rate of collision-induced thermonuclear explosions is significantly enhanced in the dense cores of GCs.

Finally, it is worth noting that the classification in \citet{2012ApJ...755..161K} allows for a diversity in the photospheric phase. The spectra of PTF09dav show similarities to those of SNe Ia, but the rest of the sample are better matched to SNe Ib/c. In the nebular phase, differences exist in the detection of oxygen emission. Nevertheless, their observed features imply comparable ejecta mass and energy release. If the ``calcium-rich'' transients are produced by one explosion channel, the model has to be able to account for this range of behaviour.

\section{Conclusions}\label{section:conclusion}
We have carried out a statistical study of the locations of ``calcium-rich''
transients within host galaxies to look for clues of progenitor properties. In
particular, this class shows strong preference for large offsets from their
apparent host galaxies, and do not follow the stellar mass distribution within
galaxies as approximated by $K$-band light. For these outlying transients,
direct measurement of properties of the local stellar population is
impossible, due to low surface brightness. Extrapolations to such large
distances are also problematic, as significant uncertainties are involved. For
our study, we choose to utilise self-consistent cosmological simulations.
We show that as a group, the ``calcium-rich'' transient locations are
consistent with relatively metal poor progenitors ([M/H] $\lesssim-0.5$).

The sample selected in \citet{2012ApJ...755..161K} includes the most extreme cases. It is not clear at the moment whether this is related to some selection bias, or sub-groups within the overall ``calcium-rich'' class. If these extreme cases are representative of a distinct class of explosions, their remote locations may imply a combination of low metallicity and old progenitor age. Events related to massive young stars are consequently disfavoured. 

The radial distribution of the ``calcium-rich'' transients is nominally
consistent with the GC distributions. However, deep non-detection limits of
the progenitors, in the few cases where they are available, disfavour an
association of the progenitors with GCs.

We expect ongoing and future surveys to generate a larger sample of transients
for statistical studies like our own, including multiple instances of newly
discovered rare types of transient events (e.g., PTF10ops).
Our method may also be used to quantify the dependence of the properties of
normal core-collapse SNe on the metallicity of their progenitors.

\section*{Acknowledgments}
This research was conducted by the Australian Research Council Centre of Excellence for All-sky Astrophysics (CAASTRO), through project number CE110001020 and through LF0992131. This research has made use of the NASA/IPAC Extragalactic Database (NED) which is operated by the Jet Propulsion Laboratory, California Institute of Technology, under contract with the National Aeronautics and Space Administration. We acknowledge use of the HyperLeda data base (http://leda.univ-lyon1.fr). This publication makes use of data products from the Two Micron All Sky Survey, which is a joint project of the University of Massachusetts and the Infrared Processing and Analysis Center/California Institute of Technology, funded by the National Aeronautics and Space Administration and the National Science Foundation.

\label{lastpage}

\end{document}